\documentclass[12pt]{article}

\usepackage{mathrsfs}
\usepackage[T1]{fontenc}
\usepackage{mathpazo}
\usepackage{setspace}
\usepackage{amsfonts}
\usepackage{amssymb}
\usepackage{amsmath}
\usepackage{epsfig}
\usepackage{latexsym}
\usepackage{color}
\usepackage{graphicx}
\usepackage{nicefrac}
\usepackage[latin1]{inputenc}
\usepackage{pstricks}
\usepackage{slashed}
\newcommand{\bea}{\begin{eqnarray}}
\newcommand{\eea}{\end{eqnarray}}
\newcommand{\be}{\begin{equation}}
\newcommand{\ee}{\end{equation}}
\newcommand{\beq}{\begin{equation}}
\newcommand{\eeq}[1]{\label{#1}\end{equation}}
\newcommand{\ber}{\begin{eqnarray}}
\newcommand{\eer}[1]{\label{#1}\end{eqnarray}}

\def\hybrid{\topmargin -20pt    \oddsidemargin 0pt
        \headheight 0pt \headsep 0pt
        \textwidth 6.25in       % A4 paper
        \textheight 9.5in       % A4 paper
        \marginparwidth .875in
        \parskip 5pt plus 1pt   \jot = 1.5ex}

\hybrid

\catcode`\@=11

\def\marginnote#1{}
\newcount\hour
\newcount\minute
\newtoks\amorpm
\hour=\time\divide\hour by60 \minute=\time{\multiply\hour by60
\global\advance\minute by-\hour}
\edef\standardtime{{\ifnum\hour<12 \global\amorpm={am}%
        \else\global\amorpm={pm}\advance\hour by-12 \fi
        \ifnum\hour=0 \hour=12 \fi
        \number\hour:\ifnum\minute<10 0\fi\number\minute\the\amorpm}}
\edef\militarytime{\number\hour:\ifnum\minute<10
0\fi\number\minute}
\def\draftlabel#1{{\@bsphack\if@filesw {\let\thepage\relax
   \xdef\@gtempa{\write\@auxout{\string
      \newlabel{#1}{{\@currentlabel}{\thepage}}}}}\@gtempa
   \if@nobreak \ifvmode\nobreak\fi\fi\fi\@esphack}
        \gdef\@eqnlabel{#1}}
\def\@eqnlabel{}
\def\@vacuum{}
\def\draftmarginnote#1{\marginpar{\raggedright\scriptsize\tt#1}}

\def\draft{\oddsidemargin -.5truein
        \def\@oddfoot{\sl preliminary draft \hfil
        \rm\thepage\hfil\sl\today\quad\militarytime}
        \let\@evenfoot\@oddfoot \overfullrule 3pt
        \let\label=\draftlabel
        \let\marginnote=\draftmarginnote
   \def\@eqnnum{(\theequation)\rlap{\kern\marginparsep\tt\@eqnlabel}%
\global\let\@eqnlabel\@vacuum}  }

\def\preprint{\twocolumn\sloppy\flushbottom\parindent 2em
        \leftmargini 2em\leftmarginv .5em\leftmarginvi .5em
        \oddsidemargin -.5in    \evensidemargin -.5in
        \columnsep .4in \footheight 0pt
        \textwidth 10.in        \topmargin  -.4in
        \headheight 12pt \topskip .4in
        \textheight 6.9in \footskip 0pt
        \def\@oddhead{\thepage\hfil\addtocounter{page}{1}\thepage}
        \let\@evenhead\@oddhead \def\@oddfoot{} \def\@evenfoot{} }

\def\numberbysection{\@addtoreset{equation}{section}
        \def\theequation{\thesection.\arabic{equation}}}

\def\underline#1{\relax\ifmmode\@@underline#1\else
        $\@@underline{\hbox{#1}}$\relax\fi}

\def\titlepage{\@restonecolfalse\if@twocolumn\@restonecoltrue\onecolumn
     \else \newpage \fi \thispagestyle{empty}\c@page\z@
        \def\thefootnote{\fnsymbol{footnote}} }

\def\endtitlepage{\if@restonecol\twocolumn \else \newpage \fi
        \def\thefootnote{\arabic{footnote}}
        \setcounter{footnote}{0}}  %\c@footnote\z@ }

\catcode`@=12 \relax

\def\figcap{\section*{Figure Captions\markboth
        {FIGURECAPTIONS}{FIGURECAPTIONS}}\list
        {Figure \arabic{enumi}:\hfill}{\settowidth\labelwidth{Figure
999:}
        \leftmargin\labelwidth
        \advance\leftmargin\labelsep\usecounter{enumi}}}
 \relax
\def\tablecap{\section*{Table Captions\markboth
        {TABLECAPTIONS}{TABLECAPTIONS}}\list
        {Table \arabic{enumi}:\hfill}{\settowidth\labelwidth{Table
999:}
        \leftmargin\labelwidth
        \advance\leftmargin\labelsep\usecounter{enumi}}}
 \relax
\def\reflist{\section*{References\markboth
        {REFLIST}{REFLIST}}\list
        {[\arabic{enumi}]\hfill}{\settowidth\labelwidth{[999]}
        \leftmargin\labelwidth
        \advance\leftmargin\labelsep\usecounter{enumi}}}
 \relax
\makeatletter
\newcounter{pubctr}
\def\publist{\@ifnextchar[{\@publist}{\@@publist}}
\def\@publist[#1]{\list
        {[\arabic{pubctr}]\hfill}{\settowidth\labelwidth{[999]}
        \leftmargin\labelwidth
        \advance\leftmargin\labelsep
        \@nmbrlisttrue\def\@listctr{pubctr}
        \setcounter{pubctr}{#1}\addtocounter{pubctr}{-1}}}
\def\@@publist{\list
        {[\arabic{pubctr}]\hfill}{\settowidth\labelwidth{[999]}
        \leftmargin\labelwidth
        \advance\leftmargin\labelsep
        \@nmbrlisttrue\def\@listctr{pubctr}}}
 \relax
\makeatother
\newskip\humongous \humongous=0pt plus 1000pt minus 1000pt

\newif\ifdtup

\relax

\def\be{\begin{equation}}
\def\ee{\end{equation}}
\def\ba{\begin{eqnarray}}
\def\ea{\end{eqnarray}}

\renewcommand{\theequation}{\arabic{equation}}
\author{
  \begin{minipage}{.97\linewidth}
    \vspace{0cm}
    \begin{center}
      \begin{small}
        \textbf{Dieter L\"ust}\footnote{dieter.luest@lmu.de} ${\ }^{1,2}$ and
         \textbf{Marios Petropoulos}\footnote{marios@cpht.polytechnique.fr} ${\ }^3$
      \end{small}
    \end{center}
    \vspace{0.5cm}
    \hspace{2cm}\begin{minipage}{.7\linewidth}
     {\it \begin{footnotesize}
    \begin{itemize}
               \item[${}^1$] Max-Planck-Institut f\"ur Physik\\
       F\"ohringer Ring 6, 80805 M\"unchen, Germany
                    \item[${}^2$] Arnold-Sommerfeld-Center f\"ur Theoretische Physik\\
        Department f\"ur Physik, Ludwig-Maximilians-Universit\"at M\"unchen\\
        Theresienstra\ss e 37, 80333 M\"unchen, Germany
            \item[${}^3$] Centre de Physique Th\'eorique\\
        Ecole Polytechnique,  CNRS UMR 7644\\
        91128 Palaiseau Cedex, France
        \end{itemize}
     \end{footnotesize}}
    \end{minipage}
    \vspace{0.5cm}
  \end{minipage}
}

\date{\today}

\title{\vspace{0.5cm}
 \boldmath \begin{large}
    \textbf{Comment on superluminality in general relativity}
  \end{large} \unboldmath
}

\begin{document}

\renewcommand{\thepage}{\arabic{page}}
\setcounter{page}{1}
%THIS IS PAGE 1 (INSERT TEXT OF REPORT HERE)

%%%%%%%%%%%%%%%%%%%%%%%%%%%%%%%%%%%%%%%%%%%%%%%%%%%%%

\begin{titlepage}
  \maketitle
  \thispagestyle{empty}

  \vspace{-13.5cm}
  \begin{flushright}
    CPHT-RR066.0911\\
  MPP-2011-115\\
    LMU-ASC 45/11
  \end{flushright}

  \vspace{12cm}

  \begin{center}
    \textsc{Abstract}\\
  \end{center}
General relativity provides an appropriate framework for addressing the issue of sub- or superluminality as an apparent effect. Even though a massless particle travels on the light cone, its average velocity over a  finite  path measured by different observers is not necessarily equal to the velocity of light, as a consequence of the time dilation or contraction in gravitational fields. This phenomenon occurs in \emph{either direction}  (increase or depletion) irrespectively of the details and strength of the gravitational interaction. Hence, it does not intrinsically guarantee superluminality, even when the gravitational field is reinforced.

\end{titlepage}

%\vskip1cm

%\tableofcontents

%\newpage

In special relativity massless particles in vacuum travel at the speed of light, $c$. This is also true in general relativity, although, in that latter case, curvature can be the source of peculiar effects affecting time, and therefore velocities. The gravitational red/blueshift, the time delay of radar echoes and, more
recently, the handling of the GPS\footnote{Global Positioning System.} clocks, provide spectacular confirmations of these properties at cosmological, solar and earth scales. 

In view of the discussions triggered by the recent OPERA data release \cite{:2011zb}, it seems appropriate to us to remind some simple facts that one should bear in mind in the process of fitting experimental data with phenomenological proposals like
via a new, environmental spin-two force that acts predominantly on neutrinos \cite{Dvali:2011mn}, or 
via warping of extra dimensions \cite{Gubser:2011mp}, or via a new scalar field sourced by the earth \cite{Kehagias:2011cb}. As it is well known, in a locally Lorentz-invariant theory, the velocity of a massless particle is always equal to $c$, in local frames. 
What we would like, however, to stress here is that, as a direct consequence of the above considerations on time,  the average velocity of such 
a massless particle along its trajectory, as measured by some observer, is not necessarily equal to $c$. It can be either smaller, or larger depending on where the measurement is performed, irrespectively of the strength and details of the gravitational field and without violating any fundamental principle  {such as local Lorentz invariance or equivalence principle}.

In order to make our argument concrete, we consider static, spherically symmetric space--times, well adapted to earth-based considerations. In a standard coordinate system, the corresponding metric reads:
\begin{equation}\label{0}
  \mathrm{d}s^2 = g_{tt}(r)
  c^2 \mathrm{d}t^2 +g_{rr}(r)\mathrm{d}r^2 +
  r^2\left(\mathrm{d}\vartheta^2 + \sin^2\vartheta\, \mathrm{d}\varphi^2\right).
\end{equation}
We perform the following \emph{gedanken} experiment: a massless particle (photon or slightly massive but very energetic particle like an electron or a neutrino) is emitted at time $t_1$ from a observer  $\mathcal{O}_1(r_1, \varphi_1, \vartheta_1)$ and received at time $t_2$ by an observer $\mathcal{O}_2(r_2, \varphi_2, \vartheta_2)$. What is the travel time measured by each of these observers and what is the corresponding average velocity?

The massless particle travels on the light cone, and we will use spherical symmetry to set $\vartheta_1=\vartheta_2=\nicefrac{\pi}{2}$. Hence, for the particle at hand
\begin{equation}\label{1}
  0= g_{tt}
  c^2 \mathrm{d}t^2 +g_{rr}\mathrm{d}r^2 +
  r^2 \mathrm{d}\varphi^2.
\end{equation}
Conservation of energy ($E$) and of angular momentum ($L$) together with Eq. (\ref{1}) allow to recast the problem in terms of a unique first-order equation,  
\begin{equation}\label{2}
 \frac{\mathrm{d}\varphi}{\mathrm{d}r}=\pm\frac{\sqrt{- g_{tt} g_{rr}}}{r\sqrt{\frac{r^2}{b^2} +g_{tt}}} ,
\end{equation}
where $b=\nicefrac{Lc}{E}$ is the impact parameter related to the periastron $R$ by setting $\left.\frac{\mathrm{d}r}{\mathrm{d}\varphi}\right\vert_R=0$, which leads to $b=\nicefrac{R}{\sqrt{- g_{tt}(R)}}$.

The total coordinate time for the flight from $\mathcal{O}_1$ to $\mathcal{O}_2$ is obtained from Eq. (\ref{1}),
\begin{equation}\label{3}
\Delta t = \frac{1}{c}\int_{r_1}^{r_2} \mathrm{d}r\sqrt{\frac{g_{rr}+r^2\left( \frac{\mathrm{d}\varphi}{\mathrm{d}r}\right)^2 }{- g_{tt}}},
\end{equation}
recast, using  (\ref{2}), as:
\begin{equation}\label{4}
\Delta t = \frac{1}{c}\int_{r_1}^{r_2} \mathrm{d}r \sqrt{\frac{ -\nicefrac{g_{rr}(r)} { g_{tt}(r)}}{1-
\frac{g_{tt}(r)}{g_{tt}(R)}\frac{R^2}{r^2}
}}.
\end{equation}
The actual time measured by the observers $\mathcal{O}_{1,2}$ at rest is, however, slightly different, because of the effect of the local gravitational potential that alters their proper time~$\tau$.
In terms of the latter one finds:
\begin{equation}\label{5}
\Delta \tau\left(\mathcal{O}_i\right) = \frac{\sqrt{-g_{tt}(r_i)}}{c}\int_{r_1}^{r_2} \mathrm{d}r \sqrt{\frac{ -\nicefrac{g_{rr}(r)} { g_{tt}(r)}}{1-
\frac{g_{tt}(r)}{g_{tt}(R)}\frac{R^2}{r^2}
}}.
\end{equation}

Expression (\ref{5}) can be applied to a large palette of situations, either for radial trajectories ($R=0$) or for more general ones. It was in particular instrumental for setting the Shapiro effect (measured in 1967 -- see e.g. \cite{Weinberg} for details and references), where, using the Schwarzschild metric  ($g_{tt}= -1+\nicefrac{r_\mathrm{g}}{r}$) and assuming $r_\mathrm{g}\ll r_1=R<r_2 $, one obtains:
\begin{eqnarray}
c\Delta \tau\left(\mathcal{O}_i\right)& =& \sqrt{r^2_2-r_1^2}-\frac{r_\mathrm{g}}{2}\left(\sqrt{\frac{r_2^2-r_1^2}{r_i^2}}-\sqrt{\frac{r_2-r_1}{r_2+r_1}}\right)\nonumber\\
&&+r_\mathrm{g}\log{\frac{r_2+\sqrt{r_2^2-r_1^2}}{r_1}} +\mathrm{O}\left(r_\mathrm{g}^2\right).
\label{6}
\end{eqnarray}
Notice that the above expressions are general enough to accommodate exotic situations\footnote{For a more conventional manifestation of repulsive gravitational forces, one can introduce a cosmological constant $\Lambda$ such that $g_{tt}= -1+\nicefrac{\Lambda r^2}{3}+\nicefrac{r_\mathrm{g}}{r}$. Equation (\ref{5}) still holds but (\ref{6}) should be adapted.} such as
a field with negative $r_\mathrm{g}$, which would result in a repulsive \emph{antigravity} potential  $\Phi=\nicefrac{MG}{r}$.

In order to determine the average speed measured by each of these observers, one has to set the distance spanned by the particle during its flight from $\mathcal{O}_1$ to $\mathcal{O}_2$. For any observer at rest, this is the length of the track line of the particle in the three-dimensional space\footnote{This definition of the length is adapted to the \emph{gedanken} experiment we are describing. Practical measurements of large distances, as they are performed in spatial physics, involve signal exchanges. Applying these methods would unnecessarily complicate our argument.}  i.e. 
\begin{equation}\label{7}
\Delta \ell = \int_{\mathcal{O}_1}^{\mathcal{O}_2} 
 \mathrm{d}\ell,
\end{equation}
where
\begin{equation}\label{8}
 \mathrm{d}\ell^2= g_{rr}(r)\mathrm{d}r^2 + r^2\left(\mathrm{d}\vartheta^2 + \sin^2\vartheta\, \mathrm{d}\varphi^2\right)
\end{equation}
with $\varphi=\varphi(r)$ the integral of Eq. (\ref{2}). 
Using the latter, (\ref{7}) reads: 
\begin{equation}\label{9}
\Delta \ell = \int_{r_1}^{r_2} \mathrm{d}r \sqrt{\frac{ g_{rr}(r)}{1-
\frac{g_{tt}(r)}{g_{tt}(R)}\frac{R^2}{r^2}
}}.
\end{equation}
Expressions (\ref{5}) and (\ref{9}) lead finally to the velocity, as measured by each of the two observers: 
\begin{equation}\label{10}
v_i=\frac{\Delta \ell }{\Delta \tau\left(\mathcal{O}_i\right)} = c\frac{\int_{r_1}^{r_2} \mathrm{d}r \sqrt{\frac{ g_{rr}(r)}{1-
\frac{g_{tt}(r)}{g_{tt}(R)}\frac{R^2}{r^2}
}}}{\sqrt{-g_{tt}(r_i)}\int_{r_1}^{r_2} \mathrm{d}r \sqrt{\frac{ -\nicefrac{g_{rr}(r)}{g_{tt}(r)}}{1-
\frac{g_{tt}(r)}{g_{tt}(R)}\frac{R^2}{r^2}
}}}.
\end{equation}

In flat space, $v_i$ is equal to $c$. However,  this no longer holds in the presence of a gravitational field,  where $v$ can be slightly \emph{above or below} $c$, depending on $g_{tt}, g_{rr}$ and on the location of the observer. Again, for Schwarzschild and under the above assumptions one finds:
\begin{equation}\label{11}
\delta_i = \frac{v_i-c}{c}= \frac{1}{2}\left(\frac{r_\mathrm{g}}{r_i}
-
\frac{r_\mathrm{g}}{
 \sqrt{r^2_2-r_1^2}
}\log{\frac{r_2+\sqrt{r_2^2-r_1^2}}{r_1}} 
\right) +\mathrm{O}\left(r_\mathrm{g}^2\right).
\end{equation}
It is clear from this expression that the average velocity can be sub- or superluminal, as a result of the competition between the two  terms of order $r_\mathrm{g}$ -- even though this effect is  illusive because the local velocity for these particles is \emph{always} identical to $c$ due to local Lorentz invariance. Hence,  $v_2<c$, whereas $v_1$ is generally superluminal.
As an example, one can consider the gravitational field around the earth ($r_\mathrm{g}\approx 1\, \mathrm{cm}$), with observer $\mathcal{O}_1$ at the surface of the earth ($r_1\approx 6.5\times10^8 \, \mathrm{cm}$) exchanging tangentially massless particles with $\mathcal{O}_2$ on (\romannumeral1) the top of the Mont Blanc ($r_2\approx 6.505\times 10^8 \, \mathrm{cm}$), (\romannumeral2) a geosynchronous satellite ($r_2\approx 4.2\times 10^9 \, \mathrm{cm}$) and (\romannumeral3) the moon ($r_2\approx 4\times 10^{10} \, \mathrm{cm}$). One finds: (\romannumeral1) $\delta_1\approx 2 \times 10^{-13} ,\delta_2\approx-4 \times10^{-13}$,  (\romannumeral2) $\delta_1\approx 4.6\times 10^{-10},\delta_2\approx-1.9  \times 10^{-10}$ and  (\romannumeral3) $\delta_1\approx 7.1 \times 10^{-10},\delta_2\approx-4.8  \times 10^{-11}$.

The above considerations do not aim at providing an explanation, based on pure general-relativity effects, to the claimed superluminal velocity of neutrinos by the OPERA collaboration \cite{:2011zb}.
Indeed, as already noticed {in \cite{Kehagias:2011cb}}, the gravitational field around the earth is too tiny to accommodate such deviations: the orders of magnitude are \emph{at best} of  $\nicefrac{r_{\mathrm{g}}}{r_{\rm earth}}\approx 10^{-9}$ and, according to our precise computation (see Eq. (\ref{11})), by far smaller when the flight distances are short -- which is the case in the OPERA experiment. 

The message our careful treatment is meant to convey, however, is that the quantity defined in general relativity as the average velocity of massless particles traveling between two distant points, can either be sub- or superluminal, depending on the position of the observer and on 
the form of the trajectory in the gravitational field. In particular, neither conventional attractive gravitational forces do systematically produce  subluminal effects, nor repulsive forces lead necessarily to superluminality. 

In our opinion, superluminality as a consequence of a (so far purely phenomenological)  modification of the effective space--time metric seen by the neutrinos\footnote{Another potential drawback of this kind of assumption is the substantial  magnification of red/blueshift effects that
a bigger (by many orders of magnitude) effective gravitational coupling might induce for these particles.} may equally originate 
(or fail to do so\footnote{Taking into account also other possible
phenomenological constraints like those originating from 5th-force experiments, and supposing the new force on the neutrinos be sourced
by the earth, seems to invalidate the attractive option  \cite{Dvali:2011mn}. However, if one gives up this assumption as well as the assumption of weak coupling, made in \cite{Dvali:2011mn}, wider scenarios with attractive forces and superluminal velocity open up
-- we thank Gia Dvali for discussions on this point.}) from an attractive or from a repulsive gravitational  force. Within this pattern, it could even be traded for subluminality, in appropriately designed experiments. Put differently, 
making the effective gravitational field for the neutrinos stronger  and repulsive does not guarantee superluminality. 
%as it was recently suggested.

%seems to be claimed in \cite{Dvali:2011mn, Kehagias:2011cb}. 

\section*{Acknowledgements}

The authors would like to thank  G. Dvali for very  fruitful correspondence as well as V. Mukhanov and T. Petkou for useful remarks.   Marios Petropoulos would like to thank the LMU for kind
hospitality. This research was supported by the Cluster of Excellence
\textsl{Origin and the Structure of the Universe} in Munich, Germany, the
French ANR contract  05-BLAN-NT09-573739, the ERC Advanced Grant  226371,
the ITN programme PITN-GA-2009-237920 and the  IFCPAR programme 4104-2. 
Finally we would also like to thank for hospitality the {\sl Paulaner} Festzelt \textsl{Winzerer Fahndl} at the Oktoberfest in Munich, where a preliminary version of this paper was set up, as well as 
%Caf\'e Puck, 
the {\sl Augustiner} Festzelt, where the final corrections were brought to completion.

%\newpage

%To be still discussed and added where and if appropriately:
%\begin{itemize}
%\item Antigravity discussion with e.g. AdS Schwarzschild $-g_{tt}= 1+k^2r^2-\nicefrac{r_\mathrm{g}}{r}$. {\bf May be this discussion is not necessary to keep the paper short. Perhaps we can introduce a parameter $\epsilon=\pm 1$ in above formulas, correspond to gravity and antigravity respectively.}
%\item Reminder: in the Newtonian limit, $g_{tt}=-1-\nicefrac{2\Phi}{c^2}$ and the Kepler problem corresponds to $\Phi=-\nicefrac{MG}{r}$. {\bf In eq.(\ref{11}), didn't you use already the Newtonioan epression?}
%\item The effect is tiny in GR to explain anything beyond the ``nanosecond'' (check the precise order of magnitude). The refined analysis shows in particular that making the gravitational field stronger or ``repulsive'' does not guarantee superluminality as it was claimed in \cite{Dvali:2011mn, Kehagias:2011cb}. Insist that the effect can go in both directions no matter what we do.{\bf See conclusions.}
%\item{Recall the more practical definition of $\Delta L$ using the exchange of photons and its consequence (valid also in the papers of our colleagues).}  {\bf ok}
%\item Our set-up is symmetric under time reversal so I found it more convenient to avoid mirrors and reflections.  {\bf ok}
%\item Perhaps we can be a bit more specific around Eq. (\ref{11}), also plugging in some number corresponding to the gravitational field of the earth. So we should compare
%the magnitudes of the first and the second terms inside the bracket.
%\end{itemize}

%I think all that has now been addressed.

\end{document}